\documentstyle[aps,prl,twocolumn,epsfig]{revtex}
\newcommand{\rd}{\,\mathrm{d}}
\newcommand{\bra}[1]{\left\langle #1 \right|}
\newcommand{\ket}[1]{\left| #1 \right\rangle}
\newcommand{\beq}[1]{\begin{equation}\label{#1}}
\newcommand{\eeq}{\end{equation}}
\newcommand{\beqar}[1]{\begin{eqnarray}\label{#1}}
\newcommand{\eeqar}{\end{eqnarray}}
\newcommand{\dd}{{\rm d}} 
\begin{document}

\preprint{TPR-99-13\\
          DUKE-TH-99-193}

\title{Angular Dependence of the Nuclear Enhancement of Drell-Yan Pairs}
\author{R.~J.~Fries$^a$, B.~M\"uller$^{a,b}$, A.~Sch{\"a}fer$^a$ and 
E. Stein$^a$}
\address{$^a$ Institut f\"ur Theoretische Physik, 
              Universit\"at Regensburg, D-93040 Regensburg, Germany\\
         $^b$ Department of Physics, Duke University, 
              Durham, North Carolina 27708-0305}

\date{\today}

\maketitle

\begin{abstract}
We calculate the nuclear enhancement in the angular distribution of 
Drell-Yan pairs produced in proton--nucleus reactions. Nuclear effects 
are encoded in universal twist-4 parton correlation functions. 
We find that the Lam-Tung relation for the angular coefficients 
of the lepton-pair distribution holds for the double-hard, but
not for the soft-hard contribution. We also predict that nuclear
enhancement effects at RHIC energies can be large.
\end{abstract}
\bigskip


The factorization theorems of QCD \cite{coll89} permit a rigorous 
treatment of hard processes by absorbing soft contributions to scattering 
matrix elements into universal, process-independent distribution 
functions.  The measurement of these distribution functions in one 
process facilitates the prediction of cross sections for other 
processes. An example is the Drell-Yan process where two hadrons 
scatter to produce a lepton pair of high invariant mass $Q^2$.
If $Q^2$ is large enough, the factorization theorems state that the 
inclusive Drell-Yan (DY) cross section can be expressed as the product 
of a calculable hard part and universal twist-2 parton distribution 
functions that can be measured, for instance, in deep-inelastic 
lepton-nucleon scattering. 

However, the structure of strong interaction dynamics is much
richer than can be described solely in terms of twist-2 structure
functions, which are two-point correlators and have a probabilistic
interpretation in terms of partons.  In reactions involving only
isolated hadrons, hard processes are usually dominated by a single 
hard scattering, and correlation functions involving more than two 
fields are normally suppressed by powers of the large scale. This
is different when reactions are considered where individual hadrons
interact with nuclei, or one nucleus interacts with another.
In that case, a  projectile parton can scatter twice within the
same nucleus of mass number $A$, and the double-scattering cross
section is enhanced by a factor $A^{1/3}$, compensating for some 
of the inherent suppression of multi-parton interactions 
\cite{lqs94,Bod89}.

Although we will focus here on proton-nucleus interactions,
our work has been primarily motivated by the fact that higher-twist 
effects should be most prominent in nucleus-nucleus collisions at 
high energy, as they will be routinely studied at the Relativistic 
Heavy Ion Collider (RHIC). Many experiments at RHIC will focus on a 
$Q^2$ range where higher-twist contributions could be significant 
\cite{HM96}. Higher-twist effects will be especially important in
the production of minijets, which are thought to provide the main
mechanism for the initial energy deposition in the central rapidity
region \cite{BM87,EKL88}.

The determination of multi-parton correlation functions in nuclei
is also an important step towards the application of QCD to nuclear 
physics, in general. The enhancement of higher-twist effects in 
reactions involving nuclei facilitates the extraction of these 
correlation functions, which encode essential aspects of the
difference in the quark-gluon structure between isolated hadrons
and nuclei. However nuclear enhancement could point to a general problem, as
it might signal an early breakdown of perturbative QCD as applied to 
high-energy collisions involving nuclei.

\begin{figure}[t]
  \centerline{\epsfig{file=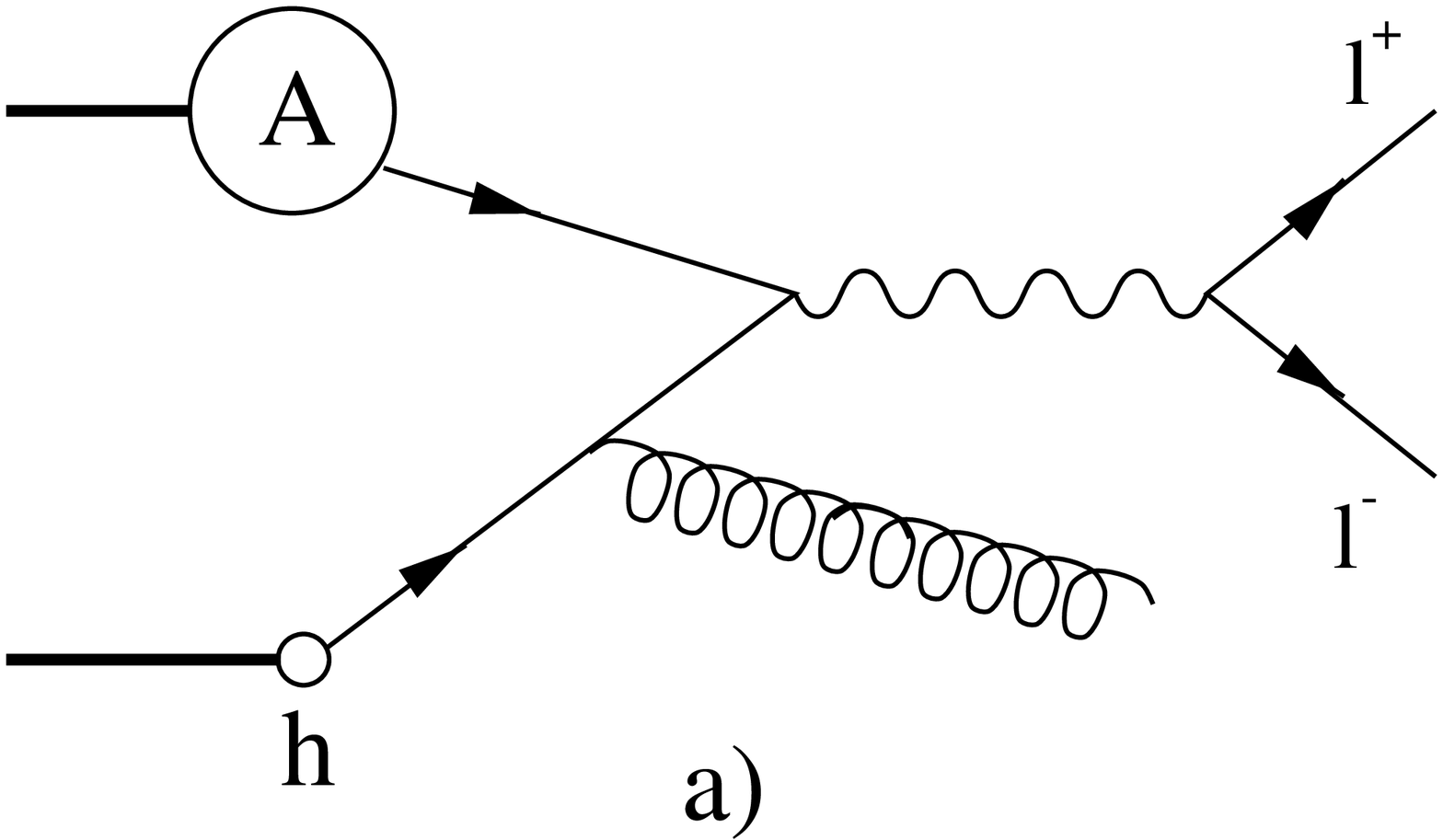,width=4.1cm} \hspace{0.2cm}
              \epsfig{file=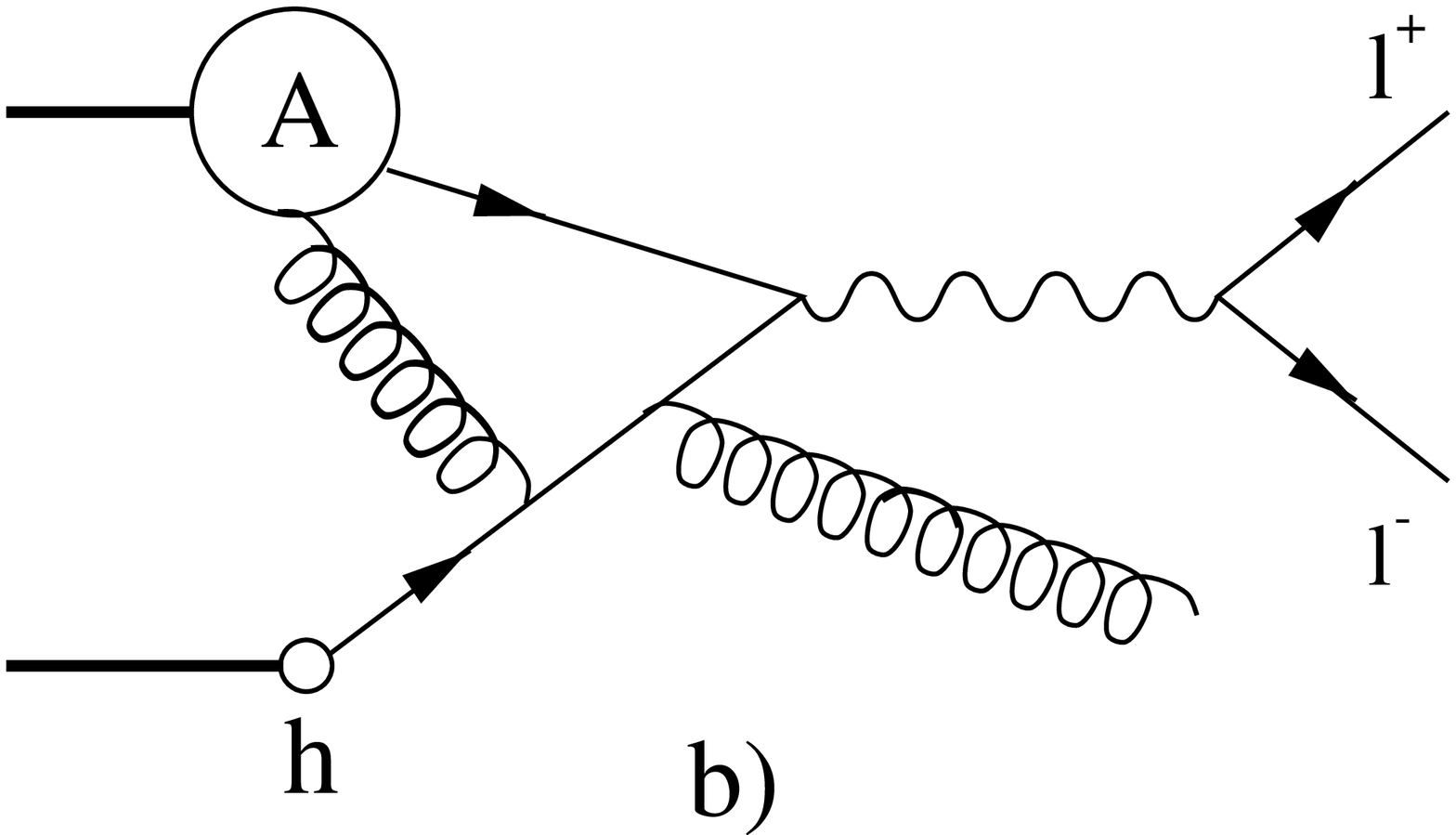,width=4.1cm}}
  \vspace{0.2cm}
  \caption{(a) Schematic single- and (b) double-scattering diagrams 
    for a hadron $h$ colliding with a nucleus $A$ and producing a lepton 
    ($l^+l^-$) pair. One additional unobserved parton is radiated with 
    high transverse momentum $q_\perp$.}
  \label{scatt}
\end{figure}

In the following we will discuss the Drell-Yan process initiated
by a single hadron $h$ scattering off a nuclear target with mass 
number $A$: 
\beq{eins}
h(P_2) + A(P_1A) \to l^+ l^- + X
\end{equation} 
At leading-twist level, a parton from the beam hadron
reacts with a single parton from the nucleus. Figure \ref{scatt}(a)
shows a typical twist-2 diagram describing production of the lepton
pair with a high transverse momentum relative to the beam axis. 
It was already pointed out some time ago by Guo \cite{guo97} that 
higher-twist effects in this process are enhanced in this kinematical
region (see also \cite{gq96}). As the parton of the beam hadron travels through
the nucleus 
it can scatter off partons from other nucleons, generating a strong 
dependence on the size of the nucleus. Such a dependence has been 
observed in several experiments \cite{e609,e683,e866} at a scale 
hard enough to resolve the partonic substructure and thus to allow 
for a treatment in the framework of perturbative QCD. 
In Fig.~\ref{scatt}(b) we give an example for double scattering.

To extract higher-twist correlation functions from experimental
data it is worthwhile to consider observables for which such
contributions are large. Ideally, one is looking for an observable 
with a vanishing contribution from single scattering.
Such quantities can be found in the angular distribution of DY pairs.
Therefore, we here extend the calculation of Ref.~\cite{guo97} to 
the differential DY cross section 
$\dd \sigma/\dd Q^2 \,\dd q_\perp^2 \,\dd y \,\dd\Omega$,
keeping the full angular dependence. In addition, we apply our
results to $p+A$ collisions at RHIC.  Some time ago, Lam and Tung 
derived a relation \cite{lt78}, similar to the Callan-Gross relation 
in deep-inelastic scattering, which states that the longitudinal 
helicity amplitude $W_L$ for the virtual photon in the DY process
is exactly twice as large as the double spin-flip amplitude 
$W_{\Delta\Delta}$. This is a leading-twist prediction valid up 
to order $\alpha_s$.

Due to the complex structure of multi-parton correlations a larger 
number of these correlation functions exist than at leading twist. 
It is useful to consider observables which receive contributions
from only a limited number of correlation functions. We find 
that the number of contributing correlation functions can be minimized 
by using different lepton-pair center-of-mass frames (see 
Ref.~\cite{lt78}). For the DY process, the so-called Gottfried-Jackson 
(GJ) frame turns out to be the most convenient choice. 
However, comparing the angular correlations
in the GJ and the Collins-Soper (CS) frames also gives valuable insight 
into the nature of double-scattering contributions \cite{cs77}.

The differential DY cross section is a classical example for a 
two-scale process in QCD, because both $Q^2$ and $q_\perp^2$, the 
transverse momentum of the photon, are detected. In order to avoid 
the occurrence of large logarithms of the type $\log^2(Q^2/q_\perp^2)$, 
which are common in this case, we require that both scales are of the 
same order. This condition also allows us to neglect interference terms 
that would introduce additional twist-4 matrix elements \cite{lqs94}.


The differential cross section for DY production is given by
\begin{equation}
\label{cross}
  d \sigma = \frac{\alpha^2}{2 S Q^4} L_{\mu\nu} W^{\mu\nu} 
  \frac{\dd^4 q}{(2 \pi)^4} \dd\Omega\, ,
\end{equation}
where $S$ is the center-of-mass energy squared, $L_{\mu\nu}$ and
$W^{\mu\nu}$ denote the leptonic and hadronic tensors,  and the angles 
$\theta$ and $\phi$ in $\dd\Omega$ refer to the polar and the azimuthal 
decay angles of the lepton pair in the rest frame of the virtual photon. 
The remaining freedom to choose these angles gives rise to the different 
choices of, e.g., the GJ and CS frames. The hadronic tensor is given by
\beq{wmunu}
W_{\mu\nu} = \int \dd^4 x \, e^{i q\cdot x} 
\bra{P_1 P_2} j_{\mu}(x) j_\nu(0)\ket{P_1 P_2} \; ,
\end{equation}
where $P_1$ and $P_2$ are the four-momenta of the nucleus (per nucleon)
and the hadron, respectively. 
The rapidity $y$ and the transverse
momentum $q_{\perp}$ of the lepton pair can be expressed in terms of the 
hadronic Mandelstam invariants 
$S = (P_1 + P_2)^2$, $T = (P_1 - q)^2$ and $U = (P_2 - q)^2$
as:
$y = \frac{1}{2} \ln( (Q^2-U)/(Q^2 - T))$
and 
$q_\perp^2 = {(Q^2-U)(Q^2 - T)}/{S} - Q^2$.

It is convenient to express the angular distribution of the lepton 
pair in terms of the helicity amplitudes $W_{\sigma,\sigma'}$ of the 
virtual photon. A complete set of such amplitudes is \cite{lt78}:
\beqar{helicity2}
W_{\rm T} &=& W_{1,1}, \quad W_{\rm L} = W_{0,0}, \quad W_{\Delta\Delta} = 
W_{1,-1} \nonumber \\
W_\Delta &=& \frac{1}{\sqrt{2}}\left(W_{1,0} + W_{0,1}\right)
\eeqar
which are referred to as the transverse, longitudinal, spin-flip, 
and double spin-flip structure functions. In terms of these helicity 
structure functions the cross section can be written as
\beqar{cross2}
\frac{\dd \sigma}{\dd Q^2 \,\dd q_\perp^2 \,\dd y \,\dd\Omega} &=&
\frac{\alpha^2}{64 \pi^3 S Q^2} \nonumber \\
&&\hspace{-1.8cm}
\left( W_{\rm TL}\left(1+\cos^2\theta\right) +
     W_{\rm L}\left(\frac{1}{2}-\frac{3}{2}\cos^2\theta\right) \right.
\nonumber \\ 
&&\hspace{-1.3cm}
+ \left. \phantom{\frac{1}{2}} W_\Delta \sin 2\theta\cos\phi +
     W_{\Delta\Delta} \sin^2\theta\cos 2\phi \right)\; ,
\eeqar
where $W_{\rm TL} = W_{\rm T} + \frac{1}{2} W_{\rm L}$.
Integration over the angles reduces the cross section to 
\beq{total}
\frac{\dd \sigma}{\dd Q^2 \,\dd q_\perp^2 \,\dd y } =
\frac{\alpha^2}{12\pi^2 S Q^2} W_{TL}\; .
\end{equation}
The ratio of (\ref{cross2}) and (\ref{total}) is often parametrized as
\begin{equation}
\label{angcoeff}
1+\lambda\cos^2\theta + \mu\sin 2\theta\,\cos\phi + 
\frac{\nu}{2}\sin^2\theta\,\cos 2\phi\; .
\end{equation}
We have verified that we reproduce the known results for the
leading-twist matrix elements \cite{cle79}.
In particular, we find $W_{\Delta\Delta} = \frac{1}{2} W_{\rm L}$, 
or $\lambda+2\nu=1$, which is the Lam-Tung relation. 


For the calculation of the double-scattering contributions we 
use the formalism of Luo, Qiu and Sterman \cite{lqs94,guo97}. 
Here we present only our main results and refer the reader to a 
forthcoming publication for details of the calculation \cite{FMSS}.
We factorize the hadronic tensor for all contributions in the
usual way, and we keep only those matrix elements, in which the two
partons from the nucleus appear separately in combinations of 
color singlet operators.  Assuming the absence of long-range color
fluctuations in the nuclear wave function, only such correlators
will show a nuclear enhancement.

It is useful to distinguish two contributions to double scattering
\cite{lqs94,guo97}. In the first case (double-hard process), 
both QCD interactions are hard and the beam parton can be considered
as on-shell between the interactions. In the second case (soft-hard
process) the beam parton picks up a soft nuclear parton before 
the hard scattering. The two contributions
differ in the pole structure of the hard part of the scattering tensor. 
We neglect the interference terms between soft and hard rescattering, 
as it is appropriate when $q_\perp$ is not too small. 
For $q_\perp^2\ll Q^2$ the interference terms are important and 
eventually spoil the nuclear enhancement \cite{guo97}.


First we investigate the case of double-hard scattering. The two-parton
distribution functions depend on two Bjorken parameters, $x_a=Q^2/(Q^2-T)$ 
and $x_h$. 
The cross section is determined by three universal nuclear matrix 
elements that were introduced in \cite{lqs94}. We only denote the
gluon-quark distribution function explicitly:
\begin{eqnarray}
\label{eqT-DH}
  T^{\rm DH}_{qg}(x_a,x_h) &=& \frac{1}{2x_h} \int \frac{{\rd} z_1^-}{2\pi} 
  \frac{{\rd} z_2^-}{2\pi} {\rd} z_3^- \Theta(z_1^- -z_2^-)
  \nonumber \\
  &&\qquad
  \Theta(-z_3^-) e^{ix_a P_1^+ z_1^-} e^{ix_h P_1^+(z_2^--z_3^-)}  
  \nonumber  \\ 
  &&\hspace{-1.2cm} 
  \bra{P_1} F^{\omega +}(z_3^-) F^{+}_{\phantom{+}\omega}(z_2^-) 
   \bar{q}(0) \gamma^+  q(z_1^-)  \ket{P_1} \; .
\end{eqnarray}
Here the superscripts $\pm$ denote the light-cone components of 
the four-vectors.
The phase factors in (\ref{eqT-DH}) restrict the integration ranges 
in such a way that one integral over the longitudinal extension of
the nucleus remains. This is the origin of nuclear enhancement.
The quark-antiquark and two-quark distribution functions, 
$T^{\rm DH}_{q\bar q}(x_a,x_h)$ and $  T^{\rm DH}_{qq}(x_a,x_h)$,
are given by similar expressions involving the correlator
$\langle \bar{q}\gamma^+  q \bar{q}\gamma^+ q\rangle$. 
The two-gluon correlator does not contribute to double-hard
scattering.

The cross section
is given by a convolution of a twist-2 structure function $f_{c/H}(\xi)$ of
the hadron, the double hard matrix element and a perturbatively 
calculable hard part $H^{ab+c} (x_a,x_h,\xi)$:
\begin{equation}
  \label{dhtensor}
  \dd \sigma \sim 
  \int \frac{\rd\xi}{\xi} 
  f_{c/H}(\xi) T^{DH}_{ab}(x_a,x_h)  H^{ab+c}(x_a,x_h,\xi).
\end{equation}
Here $a$, $b$ and $c$ denote the involved partons which have to be summed up.
Note that $x_h=x_h(\xi)$ depends on $\xi$.

We find by explicit calculation that the Lam-Tung relation holds
for double-hard scattering. In the GJ frame we also find that the 
double-hard contributions do not generate deviations from the simple 
$(1+\cos^2\theta)$ behaviour, i.e. $\mu = \nu = 0$. The explanation 
for this surprising behavior is that the double-hard process 
resembles the classical picture of double scattering. For example,
a quark from the hadron first scatters on a hard gluon from the 
nucleus, returns to its mass shell by radiating a gluon with large 
transverse momentum, and then annihilates on a hard antiquark from
the nuclear quark sea. 


For soft-hard scattering one finds that the dominant contributions
come from two universal two-parton distribution functions, 
$T^{\rm SH}_{qg}(x)$ and $T^{\rm SH}_{gg}(x)$, which depend on a 
single Bjorken parameter $x$.
They are obtained as integrals over the nuclear quark-gluon and
two-gluon correlators, 
$\langle F^{\omega +}\bar{q}\gamma^+ qF^{+}_{\phantom{+}\omega}\rangle$ 
and
$\langle F^{\omega +}F^{+}_{\phantom{+}\omega} 
F^{\lambda +}F^{+}_{\phantom{+}\lambda}\rangle$, 
respectively. Following Luo et al.~\cite{lqs94}, we
neglect the contributions from two-quark correlators involving
one soft quark, because these are far exceeded by the 
correlators involving a soft gluon.  Again, assuming the absence 
of long-range color fluctuations in the nucleus, an enhancement 
proportional to the nuclear radius is obtained. Explicit evaluation
of the soft-hard cross section reveals that the Lam-Tung relation
is violated by this contribution.

Before we present numerical results, we need to specify the nuclear 
twist-4 matrix elements. In the absence of experimental results,
we follow \cite{lqs94,mq86} and express the 
two-parton distribution functions $T^{\rm DH}$ as products of the 
known one-parton distribution functions $f(x)$:
\begin{equation}
T^{\rm DH}_{ab}(x_a,x_h) = \bar{C}(A) A\, f_{a/A}(x_a) f_{b/A}(x_h)\; .
\end{equation}
Here $f_{a/A}$ is the parton distribution (neglecting shadowing
and EMC-type effects) in a nucleus with mass number $A$.
The normalization constant is assumed to grow like the nuclear radius,
$\bar{C}(A) = CA^{1/3}$ with $C=0.072$~GeV$^2$ \cite{guo97}.
For soft-hard scattering the twist-4 matrix elements only depend 
on the momentum of the hard parton. We assume that they are 
proportional to the distribution functions of the hard parton 
and include the effect of the presence of an additional soft gluon 
as a renormalization:
\begin{equation}
T^{\rm SH}_{ab}(x) = \lambda_{\rm LQSG}^2 A^{4/3} f_{a/A}(x)
\end{equation}
where $\lambda_{\rm LQSG}=0.1$~GeV is taken from \cite{guo98}. We use the 
twist-2 distribution functions from the CTEQ3M parameterization 
for the proton \cite{cteq3}. 

\begin{figure}[t]
  \centerline{\hspace{-2cm}
              \epsfig{file=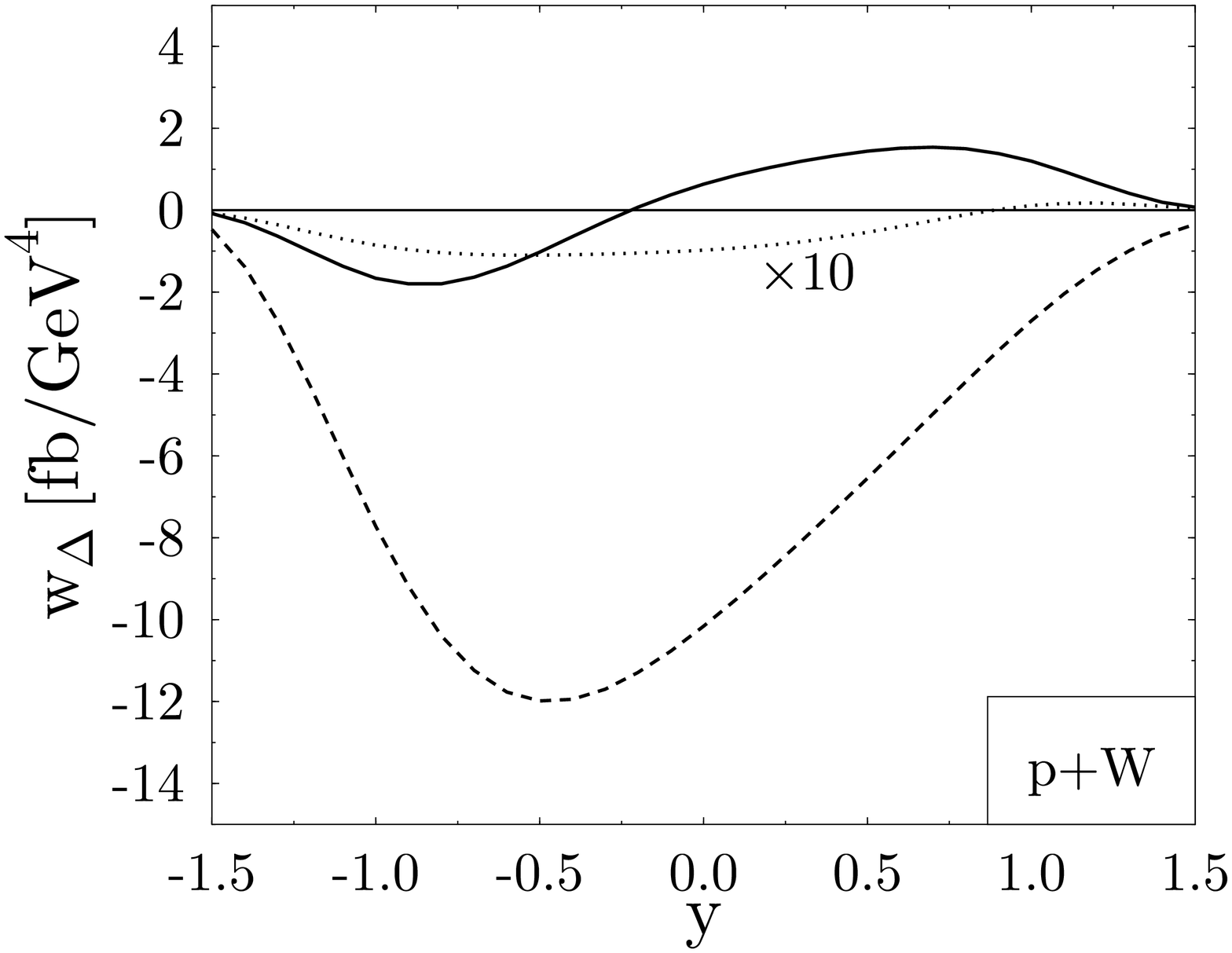,width=\columnwidth,height=4.8cm}}
  \vspace{0.2cm}
  \caption{The helicity amplitude 
    $w_\Delta = \alpha^2W_\Delta/(64\pi^3SQ^2)$ for 800 GeV $p+W$ 
    at $Q= 5$~GeV and $q_\perp=4$~GeV. The plot shows the twist-2 result
    (solid line), the double-hard (dashed line) and soft-hard 
    (dotted line) twist-4 contributions, in the CS frame. 
    The soft-hard result is 
    multiplied by a factor of 10. } 
  \label{fige866}
\end{figure}

We first discuss the application of our results to the data for
800 GeV protons on $W$ from the experiment E866 at Fermilab \cite{e866}.
Figure~\ref{fige866} shows the contribution from the single spin-flip
amplitude $W_\Delta$ to the differential DY cross section (\ref{cross2}) in 
the CS frame.
(Note that $W_\Delta$ vanishes in $p+p$ collisions for symmetry reasons.)
The twist-4 contribution is much larger than the twist-2 result, and it
is completely dominated by the double-hard scattering process.
Note that the soft-hard contribution is negligible. This feature can be
tested by checking experimentally for a violation of the Lam-Tung relation
which is exclusively given by soft-hard processes.
Furthermore a characteristic property of our soft-hard and double-hard 
contributions is the specific difference between the CS and GJ 
frames \cite{FMSS}. 

\begin{figure}[t]
  \vspace{-0.5cm}
  \centerline{\hspace{-2cm}
              \epsfig{file=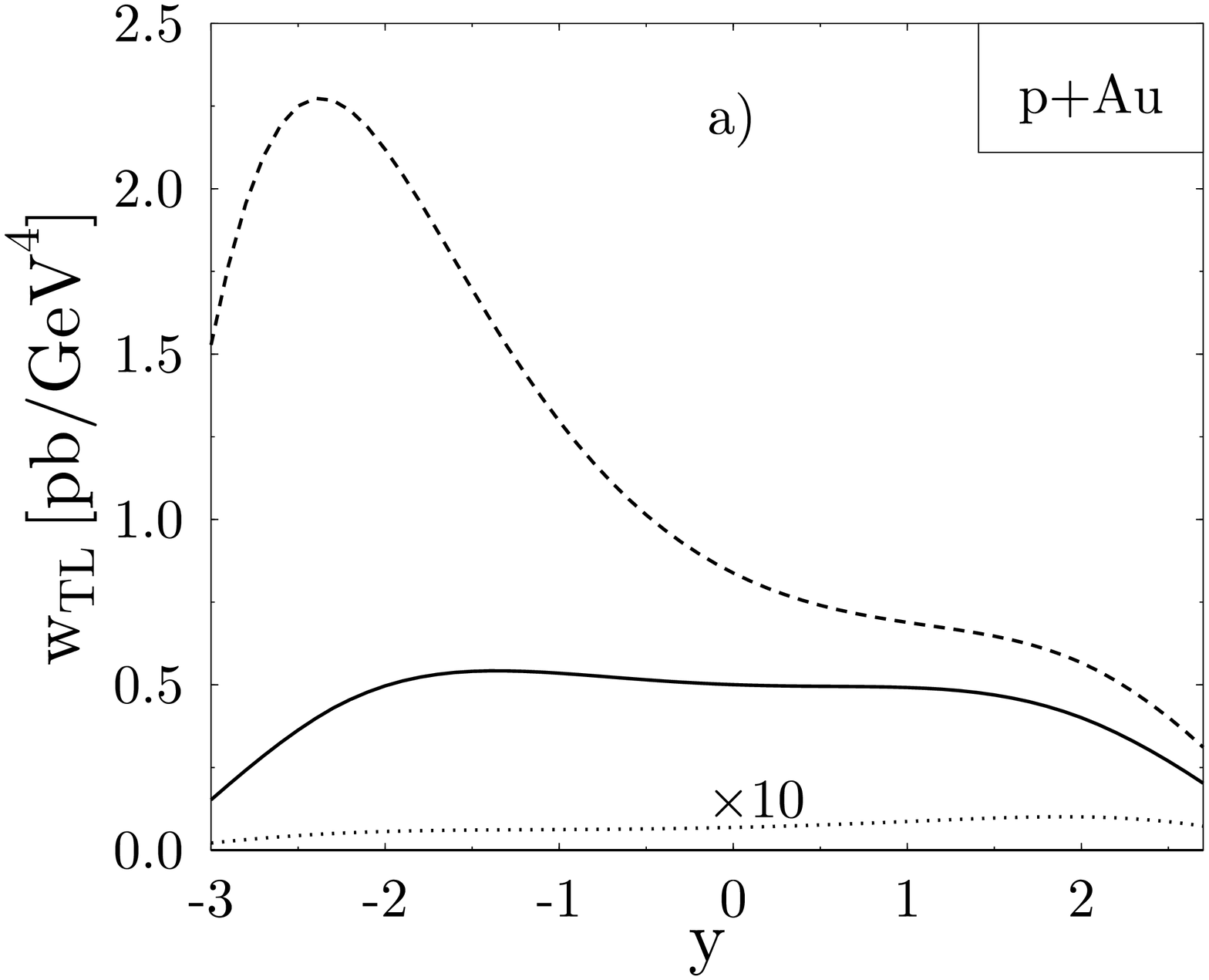,width=\columnwidth,height=4.8cm} } 
  \vspace{0.4cm}
  \centerline{\hspace{-2cm}
              \epsfig{file=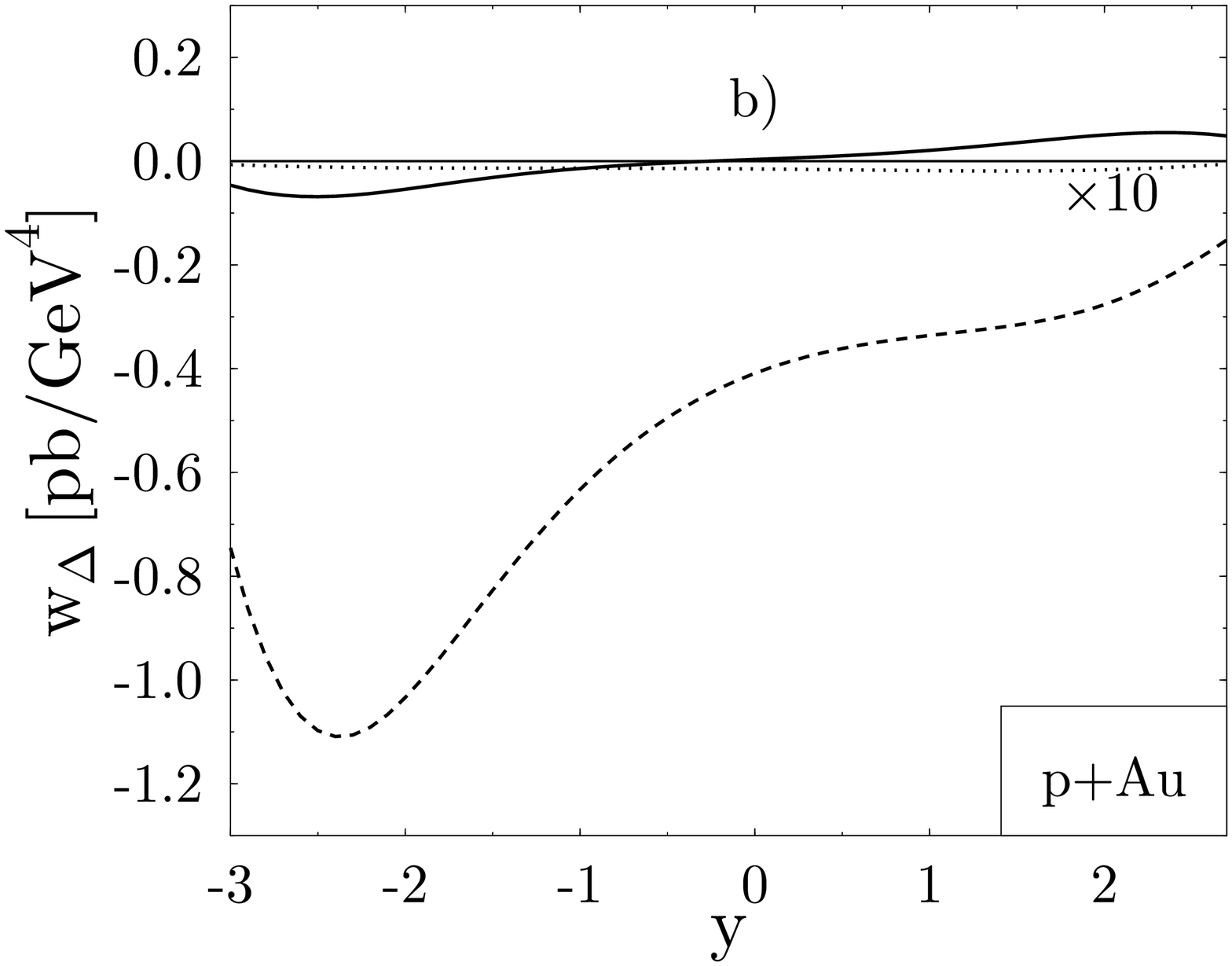,width=\columnwidth,height=4.8cm}}

  \vspace{0.2cm}
  \caption{The helicity amplitudes (a) $w_{\rm TL}$ and (b) $w_\Delta$ in the 
    CS frame for 250~GeV protons colliding with 100~GeV/nucleon $Au$ nuclei at 
    $Q=5$~GeV and $q_\perp=4$~GeV. For details see Fig.~\protect\ref{fige866}.}
  \label{figrhic}
\end{figure}

Next, we turn to $p+Au$ collisions at the full RHIC energy. As 
Fig.~\ref{figrhic} shows, the twist-4 contribution is again larger
than the leading-twist result. The twist-4 cross section peaks at
negative rapidities (the direction of the proton beam), because 
the requirement of a high transverse momentum enhances the contribution, 
in which a quark from the proton annihilates with an antiquark from the 
$Au$ nucleus. Again, the soft-hard process is negligible. The kinematics
of the peak ($Q=5$~GeV, $q_\perp=4$~GeV, $y\approx -2$) fits well with the
acceptance of the {\sc Phenix} detector \cite{phenix}. The predicted
count rate is not high (about 30 events/year in a window of 1 GeV$^4$),
but the characteristic forward-backward asymmetry should be easily
detectable.

In summary, we have calculated the nuclear enhancement of the twist-4
contribution to DY production in $p+A$ collisions, with its full
angular distribution. 
For RHIC energies the soft-hard contributions involving highly
non-trival multi-field correlators are negligible, which
is good news for any comprehensive QCD-description of high-energy
heavy ion collisions.
The dominant contribution from 
double-hard scattering has an interpretation in terms 
of the classical double scattering picture.
While double-hard scattering respects the Lam-Tung
relation, soft-hard scattering does not. Double-hard scattering
does not contribute in the GJ-frame, which allows 
for a sensitive experimental test of the LQS-formalism.
It would be very interesting to calculate the same effect
in other approaches \cite{Bod89,KTS99}.
A measurement of the nuclear enhanced contribution should 
be possible at RHIC.

{\bf Acknowledgements.}
We are grateful to X.Guo for helpful correspondence. 
We benefitted from discussions with V.~Braun, L.~Szymanowski, and O.~Teryaev. 
This work was supported in part by the BMBF and GSI. 
BM acknowledged support from the Alexander von Humboldt
Foundation 
(U.S.~Senior Scientist Award) 
and a grant from
the U.S.~Department of Energy (DE-FG02-96ER40495).


\begin{thebibliography}{99}

\bibitem{coll89}
  J.~C.~Collins, D.~E.~Soper and G.~Sterman, in: 
  {\em Perturbative Quantum Chromodynamics}, ed.  
  A.H.~Mueller (World Scientific, Singapore, 1989) 
  and references therein.

\bibitem{lqs94}
  M.~Luo, J.~Qiu and G. Sterman,
  Phys.~Rev.~{\bf D49}, 4493 (1994); 
  {\em ibid.}~{\bf D50}, 1951 (1994).  

\bibitem{Bod89}
  G.T.~Bodwin, S.J.~Brodsky and G.P.~Lepage,
  Phys. Rev. {\bf D39}, 3287 (1989); 
  R.~Baier, Y.~L.~Dokshitzer, A.~H.~Mueller, S.~Peigne, and D.~Schiff,
  Nucl.~Phys.~{\bf B484}, 265 (1997)

\bibitem{HM96}
  J.~W.~Harris and B.~M\"uller,
  Ann. Rev. Nucl. Part. Sci. {\bf 46}, 71 (1996).

\bibitem{BM87}
  J.~P.~Blaizot and A.~H.~Mueller,
  Nucl.~Phys.~{\bf B289}, 847 (1987).

\bibitem{EKL88}
  K.~J.~Eskola, K.~Kajantie, and J.~Lindfors,
  Phys.~Lett. {\bf B214}, 613 (1988).

\bibitem{guo97}
  X. Guo,
  Phys.~Rev.~{\bf D58}, 036001 (1998).

\bibitem{gq96}
  X.~Guo and J.~Qiu,
  Phys. Rev. {\bf D53}, 6144 (1996);
  X.~Guo,
  Phys. Rev. {\bf D58}, 114033 (1998)

\bibitem{e609}
  M.~D.~Corcoran et al.,
  Phys.~Lett.~{\bf B259}, 209 (1991).

\bibitem{e683}
  D.~Naples et al.,
  Phys.~Rev.~Lett.~{\bf 72}, 2341 (1994).

\bibitem{e866}
  M.~A.~Vasilev et al.,
  hep-ex/9906010.

\bibitem{lt78}
  C.~S.~Lam and Wu-Ki~Tung, 
  Phys.~Rev.~{\bf D18} 2447 (1978); 
  {\em ibid.}~{\bf D21} 2712 (1980).

\bibitem{cs77}
  J.~C.~Collins and D.~E.~Soper,
  Phys.~Rev.~{\bf D16}, 2219 (1977).

\bibitem{cle79}
  J.~Cleymans and M.~Kuroda,
  Nucl.~Phys.~{\bf B155}, 480 (1979); 
  Erratum: \emph{ibid}.~{\bf B160}, 510 (1979); 
  E.~Mirkes, 
  Nucl.~Phys.~{\bf B387}, 3 (1992) 

\bibitem{FMSS}
  R.~J.~Fries, B.~M\"uller, A.~Sch\"afer, and E.~Stein,
  manuscript in preparation.

\bibitem{mq86}
  A.~H.~Mueller and J.~Qiu,
  Nucl.~Phys.~{\bf B268}, 427 (1986).


\bibitem{guo98}
X.~Guo,
Phys. Rev. {\bf D58}, 114033 (1998)


\bibitem{cteq3}
  H.~L.~Lai et al.,
  Phys.~Rev.~{\bf D51}, 4763 (1995).

\bibitem{cs77}
  J.~C.~Collins and D.~E.~Soper,
  Phys.~Rev.~{\bf D16}, 2219 (1977).

\bibitem{phenix}
  S.~R.~Tonse and J.~H.~Thomas,
  in: {\em Pre-Equilibrium Parton Dynamics},
  ed. X.-N.~Wang, p. 341, Report LBL-34831 (1993).

\bibitem{KTS99}
B.Z.~Kopeliovich, A.V.~Tarasov and A.~Sch\"afer,
Phys. Rev. {\bf C59}, 1609 (1999) ; 
B.Z.~Kopeliovich, A.V.~Tarasov and A.~Sch\"afer,
hep-ph/9908245; 
U.A.~Wiedemann and M.~Gyulassy,
hep-ph/9906257.

\end{thebibliography}
\end{document}